\pdfoutput=1
\documentclass[nonacm,10pt,sigconf,a4paper]{acmart}
\usepackage[utf8]{inputenc}
\usepackage{subcaption}
\renewcommand\footnotetextcopyrightpermission[1]{} 

\title{A Multi-Perspective Study of Internet Performance during the COVID-19 Outbreak}

\author{Ahmed Elmokashfi}
\email{ahmed@simula.no}
\affiliation{%
  \institution{SimulaMet}
  \city{Oslo}
  \state{Norway}
}

\author{Alfred Arouna}
\email{alfred@simula.no}
\affiliation{%
  \institution{SimulaMet}
  \city{Oslo}
  \state{Norway}
}

\author{Ioana Livadariu}
\email{ioana@simula.no}
\affiliation{%
  \institution{SimulaMet}
  \city{Oslo}
  \state{Norway}
}

\author{Mah-Rukh Fida}
\email{mahrukhfida@simula.no}
\affiliation{%
  \institution{SimulaMet}
  \city{Oslo}
  \state{Norway}
}

\author{Amund Kvalbein }
\email{amundk@simula.no}
\affiliation{%
  \institution{SimulaMet}
  \city{Oslo}
  \state{Norway}
}

\author{Anas Al-Selwi}
\email{anasal@simula.no}
\affiliation{%
  \institution{SimulaMet}
  \city{Oslo}
  \state{Norway}
}

\author{Thomas Dreibholz}
\email{dreibh@simula.no}
\affiliation{%
  \institution{SimulaMet}
  \city{Oslo}
  \state{Norway}
}

\author{Haakon Bryhni}
\email{haakonbryhni@simula.no}
\affiliation{%
  \institution{SimulaMet}
  \city{Oslo}
  \state{Norway}
}

\begin{document}
\renewenvironment{comment}{}{}

\begin{abstract}
The rapid spread of the novel corona virus, SARS-CoV-2, has prompted an unprecedented response from governments across the world. 
A third of the world population have been placed in varying degrees of lockdown, and the Internet has become the primary medium for conducting most businesses and schooling activities. 
This paper aims to provide a multi-prospective account of Internet performance during the first wave of the pandemic.
We investigate the performance of the Internet control plane and data plane from a number of globally spread vantage points. 
We also look closer at two case studies. First, we look at growth in video traffic during the pandemic, using traffic logs from a global video conferencing provider.
Second, we leverage a country-wide deployment of measurement probes to assess the performance of mobile networks during the outbreak. 
We find that the lockdown has visibly impacted almost all aspects of Internet performance.  
Access networks have experienced an increase in peak and off-peak end to end latency. 
Mobile networks exhibit significant changes in download speed, while certain types of video traffic has increased by an order of magnitude.  
Despite these changes, the Internet seems to have coped reasonably well with the lockdown traffic.


\end{abstract}

\maketitle

\section{Introduction}
In the spring of 2020, a third of the world population has been placed in varying degrees of lockdown to slow down the spread of the novel corona virus, SARS-CoV-2.
Public services, businesses, schools and universities around the world have been shut down.
Virtual classrooms and video conferences quickly replaced the regular classroom and the physical office space.
The state of our economies immediately became contingent on the Internet infrastructure, further emphasising the critical role of the Internet to our livelihood.

This unprecedented situation has strained both access networks and the backbone of the Internet.
Several network operators have reported excessive growth in traffic.
For example, Telefonica's data traffic growth in March alone has overtaken its normal annual growth~\cite{eu_lockdown_covid19}. 
Industry and regulatory reports have since shown that the Internet has, in general, coped reasonably well with lockdown traffic ~\cite{google,ncta,fastly,telegeography,berec,cable-uk,ookla-two}.
However, a number of studies have also pointed to variations in performance, increase in outages and under-performing access networks~\cite{rajiullah2020mobile,Angelique-covid,candela,liu2020characterizing}  

Almost all available accounts of Internet performance during the lockdown are from the point of view of a single operator , a few operators or a few IXPs.
This paper aims to provide a broader multi-perspective account on the Internet performance performance and stability during the first wave of Covid-19.
We examine how the global routing system and the Domain Name Systems (DNS) have coped. 
In addition, we analyze latency between thousands of vantage points worldwide enabled by RIPE Atlas ~\cite{ripe_atlas} to determine the impact on performance change in Internet usage during the first wave of the pandemic.
Using traffic logs from a global network that facilitates high quality video conferencing, we examine growth in video traffic during the pandemic. 
Finally, we take a closer look at the performance of data traffic in a mobile network in a Western European country to examine the impact of the lockdown.

This approach disentangles us from the limited single-view approach and allows for examining effects beyond a simple binary classification. 
For instance, examining the routing system allows for capturing effects like the link between network stability and human intervention.
Also, this approach allows us to look at differences between different countries.
The traffic from the global video interconnect gives an idea about how businesses in different regions have coped with the lockdown and may provide an input to evaluating the economical impact. 

We find that the pandemic and the following lockdown measures have had a clear impact on the Internet. 
BGP routing dynamics in the Internet have changed during the lockdown period, with a reduction in transient routing changes and increased signs of traffic engineering. 
Increased traffic volumes have stressed the capacity limits of many networks, and we observe increased delays in most countries where lockdown measures have been imposed. 
We also see a clear increase in outages during the pandemic, with European countries being over-represented.
The performance of DNS root servers, on the other hand, does not seem to be affected.
Our case study from a global video conferencing network indicates a sharp increase in video traffic globally, with traffic levels increasing more than ten times for certain customers.
Likewise, the performance of two mobile networks in a Western European country were clearly affected, with reduced data speeds in densely populated residential areas and increased speeds outside urban areas as people travel less and stay at home.

The rest of this paper is organized with an initial discussion of impact on the control plane by analyzing BGP data in Sec.~\ref{sec:control}, followed by an analysis of impact on the data plane in Sec.~\ref{sec:Data}, outages during the outbreak in Sec.~\ref{sec:outages}, a study of the impact of DNS is provided in Sec.~\ref{sec:dns}, a case study of a video conferencing provider in Sec.{\ref{sec:video}}, and a case study using a country-wide measurement platform for mobile broadband performance before and after the pandemic in Sec.
~\ref{sec:speed}. The results are discussed and compared to related work in Sec.~\ref{sec:discussion} and Sec.~\ref{sec:Related}. Finally, Sec.~\ref{sec:Conclusion} concludes the paper.

\label{sec:Intro}


\section{Impact on control plane}
We investigate whether the lockdown had a visible impact on the interdomain routing system.
To this end, we focus on three aspects:  impact on the growth of the system, its dynamics and stability. 

\subsection{Dataset}
We use BGP routing table dumps and update traces from the Routeviews Oregon-IX collector. 
We focus on two periods, the 1st of January to the 30th of April 2019 and 2020, respectively. 
These two periods allow us to compare the same months before and during the COVID-19 outbreak. 
The chosen collector peers with up to 49 monitors from over 35 Autonomous systems (ASes).

\subsection{Impact on growth} 
We record a slower growth in the size of the default-free IPv4 routing table in the period January-April 2020 compared with the same period in 2019. 
The growth rate was 3.2\% in 2019, but it has dropped to 1.7\% in 2020. 
The number of newly appearing autonomous systems is comparable in both years, albeit a bit slower in 2020. 
While this slowness might be related to the general slowness of business during lockdown, the depletion of IPv4 address space is a possible explanation too.   
We also investigate the structure of the routing table in terms of type of prefixes. 
In particular, we divide prefixes into top, lonely, delegated and deaggregated. 
Top and Lonely prefixes are not covered by any other routed prefix.
Lonely prefixes do not cover any routed prefix, whereas top prefixes cover more specifics (i.e. longer sub-prefixes). 
Address space classified as deaggregated and delegated is covered by a less specific prefix in the routing table. 
Deaggreagated space is advertised by the same AS that advertises the less specific prefix, while delegated space is advertised by a different AS. 
Interestingly, we measure a slight drop in the fraction of both deaggregated and delegated -- about 2\% for each.
This is counter-intuitive, since one could expect an increase in deploying traffic engineering mechanisms like prefixes deaggregtion as a response to rising traffic demands.
One explanation can be that networks have resorted to other approaches for deploying traffic engineering like the use of communities.
Another possible explanation is that traffic growth has mainly impacted domestic routes.
In other words, telecommuting, homeschooling and online public services often involve exchanging data between users in the same country and even within a limited geographic scope.

\subsection{Impact on routing dynamics}
\begin{figure}[t]
\centering
\includegraphics[width =0.75\columnwidth, angle=270]{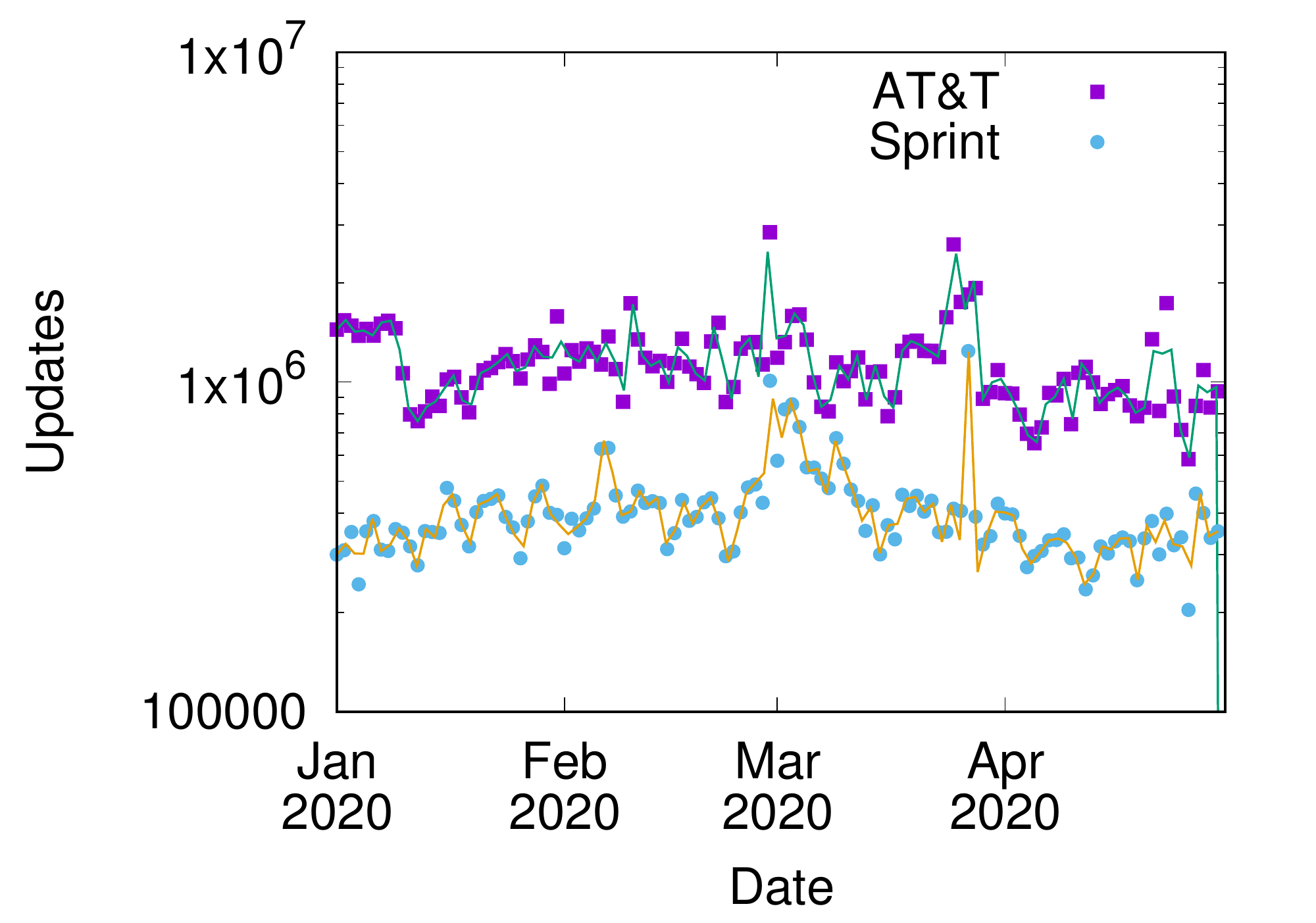}
\caption{The daily number of updates in the first four months of 2020 sent by routers in AT\&T and Sprint networks}
\label{fig:updates-ts}
\end{figure}

We also investigate the growth in BGP updates, since this gives an idea about the stability of the interdomain routing system. 
We expect, as previous work has shown~\cite{elmokashfi2013revisiting}, the number of updates to grow as a function of the routing system size.
Surprisingly, however, the visual inspection of daily updates time series show that the majority of monitors exhibit a decreasing trend.
Figure~\ref{fig:updates-ts} shows the timeseries of the total daily updates that are sent by two monitors in AT\&T and Sprint networks, respectively.
Both timeseries exhibit a drop starting at the end of March, which coincides with the period where many countries took drastic lockdown measures.

To rigorously confirm and quantify the visual observations, we apply the Mann-Kendall ({\em MK}) trend detection test to all timeseries in 2019 and 2020.
The MK test is a non-parametric test that detects monotonically linearly increasing or decreasing trends and returns the corresponding slope, which is refereed to as Sen's slope~\cite{hollander2013nonparametric}.
We accept the test result if the $p-value \leq 0.05$.

\begin{table}[h!]
\centering
\begin{tabular}{ |c|c|c|c| } 
\hline
{\bf year} & {\bf No trend} & {\bf Increasing} & {\bf Decreasing} \\
\hline
2019 & 17 (34.7\%) & 25 (51.0\%) & 7 (14.3\%)
\\
\hline
2020 & 21 (46.7\%) & 6 (13.3\%) & 18 (40.0\%)
\\
\hline
\end{tabular}
\caption{Trend in daily update time series.}
\label{tab:trend}
\end{table}

Table~\ref{tab:trend} summarizes the results of the {\em MK} test.
Unlike in 2019, 86\% of the monitors exhibit no trend or decreasing trend in 2020. 
A tiny 13.3\% show an increasing trend in 2020 compared to over half of the monitors in 2019.  
The median slope of the increasing trend in 2019 is 2082 updates/day, while the median slope of the decreasing trend in 2020 is -2105 updates/day. 
The similarity between these slopes is very interesting.   
These results indicate that the cause of daily increase in the sustained rate of BGP updates has been muted in the first few months of 2020.
Hence, resulting in either a stable number of updates or the complete reversal of the daily trend.

A closer look at the monitors with an increasing trend reveals that half of them have experienced an upward level shift in the number of daily updates, which is interpreted as an increasing trend. 
Previous work have shown that level shifts in BGP updates often coincide with effects local to the monitors e.g. misconfigurations ~\cite{elmokashfi2011bgp}. 
Notably, over half of the monitors with a decreasing trend in 2019 exhibit a downward level shift.
We conjecture, accordingly, that both the decreasing trend in 2019 and increasing trend in 2020 reflect effects local to the respective monitors.

\begin{figure}[t]
\centering
\includegraphics[width =0.75\columnwidth, angle=270]{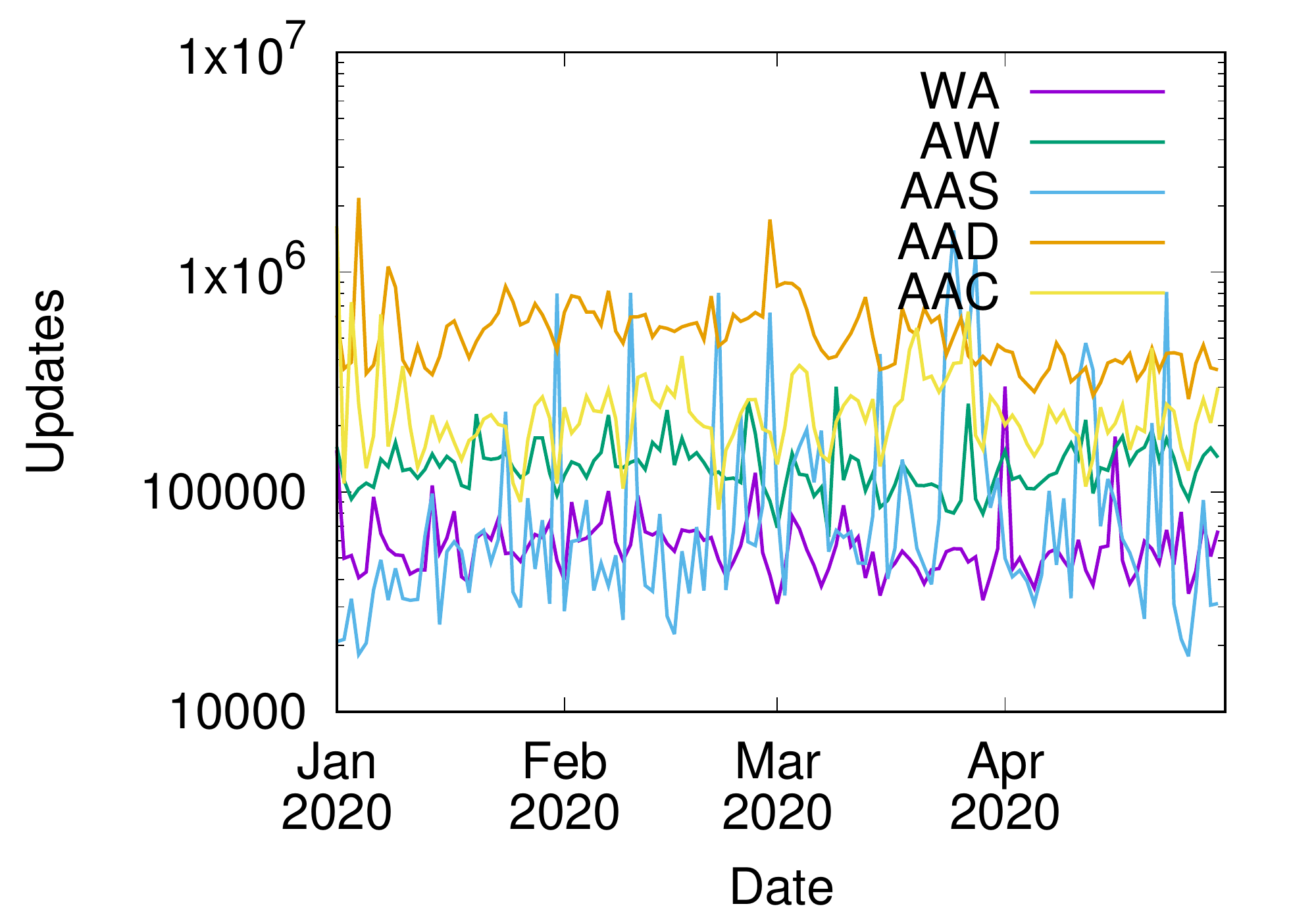}
\caption{Updates split per routing change type for the AT\&T monitor.}
\label{fig:evts-ts}
\end{figure}

\subsection{Impact on routing changes}

In order to narrow down the possible causes behind the decrease in daily BGP updates, We investigate the underlying routing changes.
To this end, we group routing updates that arrive close in time into routing events.
As in ~\cite{wu2005finding}, we consider consecutive updates, concerning a routing prefix $p$, that are spaced by no more than 70 seconds as part of the same underlying routing change.
More specifically, we start by scanning the timeseries of updates and once we encounter an update for a prefix $p$, we flag it as a start of a routing event.
The next update is considered as part of the same event, if it is no longer than 70 seconds out. 
Otherwise, we label the previous event as finished and start a new event.
The 70 seconds thresholds is chosen to account for the effect of the MRAI timer that delays BGP updates transmission. 

We next classify routing events according to the starting $PA_{st}$ and ending $PA_{en}$ AS paths into: 1) {\bf AAC}: $PA_{en}$ is different from $PA_{st}$, 2) {\bf AAD}: $PA_{en}$ is the same as $PA_{st}$ and we see at least one different AS Path during the event, 3) {\bf AAS}: $PA_{en}$ is the same as $PA_{st}$ and we do not see any different AS Path during the event. $PA_{en}$ and $PA_{st}$ may have different path attributes (e.g. communities and MED values), 4) {\bf AW}: the event ends by withdrawing the affected prefix from the routing table i.e., $PA_{en} = \emptyset$, 5) {\bf WA}: the event announces a new or a previously withdrawn prefix i.e., $PA_{st} = \emptyset$.      

Figure~\ref{fig:evts-ts} depicts the breakdown of daily updates by event type for a monitor in AT\&T network. 
The updates related to the AAD events have dropped starting early March, while those related to AAC and AAS have either slightly risen or became more erratic.
The contributions of AW and WA events have remained stable.
The above observations apply to most monitors that exhibit a decreasing trend in daily updates.
This indicates that most decrease is related to the drop in AAD changes.
Since these changes dominate the mix of updates, the increase in AAC and AAS changes does not seem to offset the decrease.

\begin{figure}[ht]
\begin{subfigure}{0.4\textwidth}
\includegraphics[width=0.6\columnwidth, angle=270]{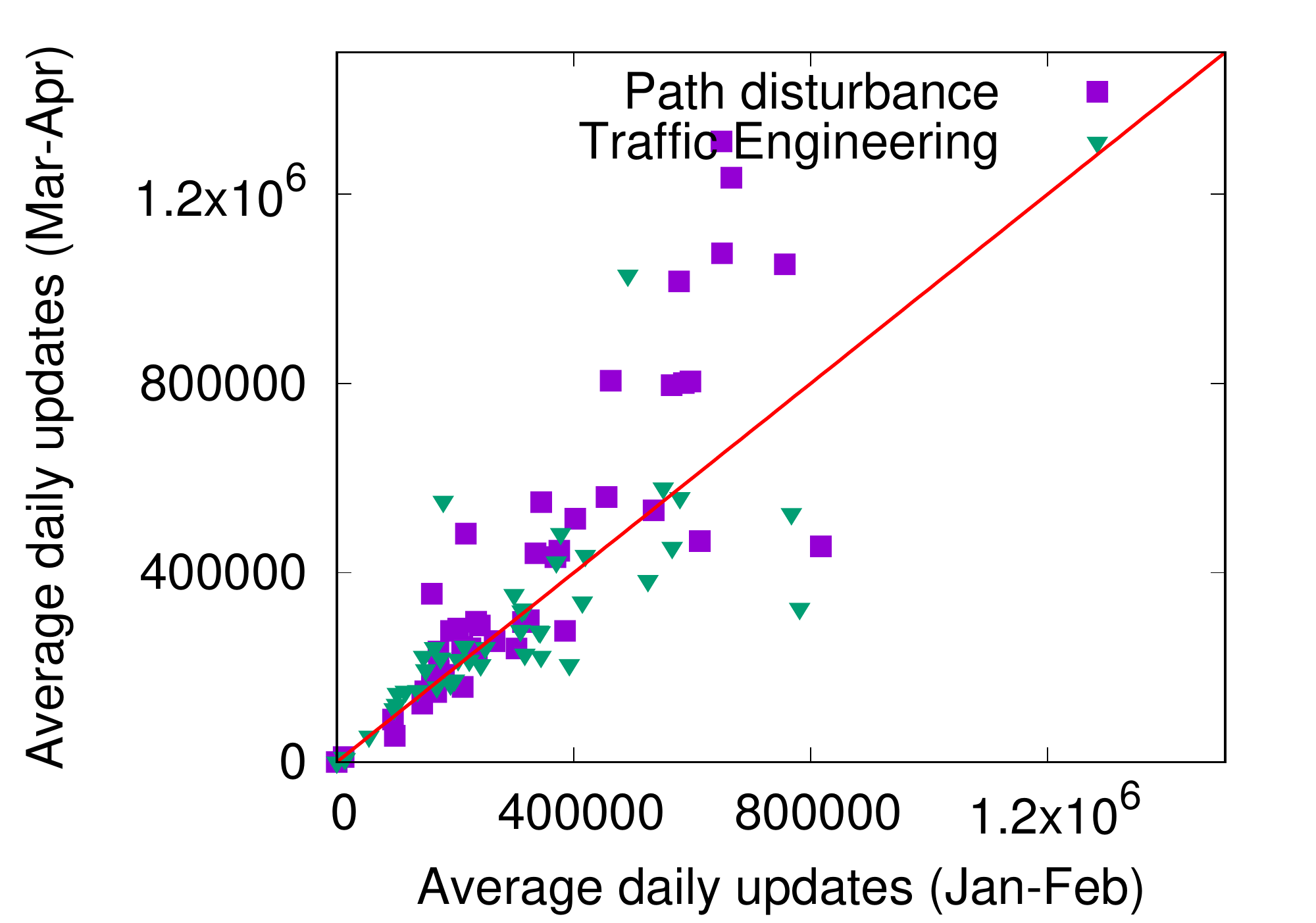} 
\caption{2019}
\label{fig:se-2019}
\end{subfigure} 
\begin{subfigure}{0.4\textwidth}
\includegraphics[width=0.6\columnwidth, angle=270]{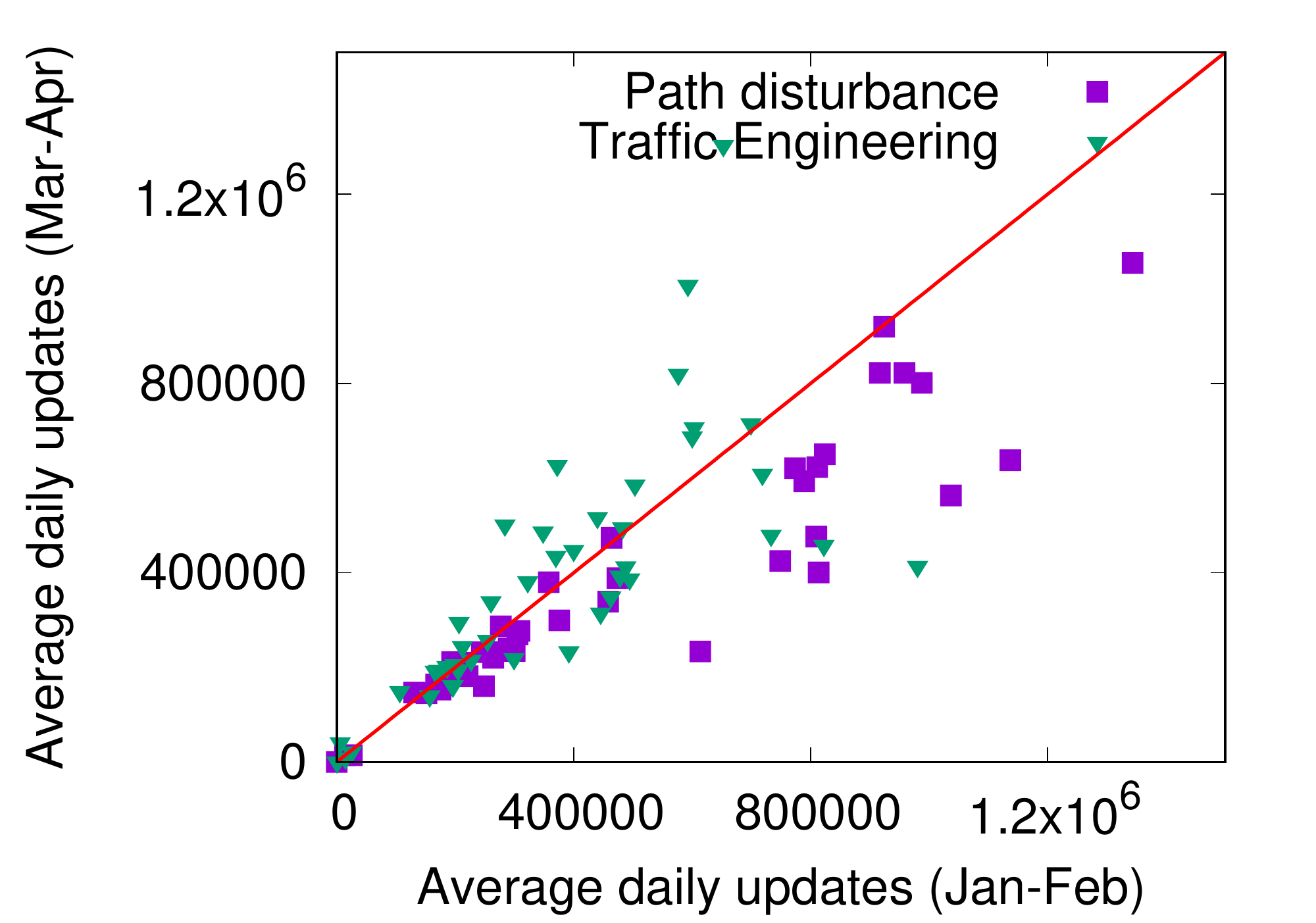}
\caption{2020}
\label{fig:se-2020}
\end{subfigure}
\caption{Routing changes comparison}
\label{fig:scatter-event}
\end{figure}

AAD changes communicate routing disturbances that are likely caused by transient routing changes e.g. router restarts or re-configurations. 
AAC changes reflect longer term routing changes, while AAS reflects configuration changes within the origin AS like changing communities.
The latter two are often used to influence traffic, i.e. traffic engineering.  
The two panels in Fig.~\ref{fig:scatter-event} are scatter plots of the average daily updates related to path disturbance and traffic engineering, for 2019 and 2020, respectively. 
We plot the daily averages for the period January-February on the x-axis and daily averages for the period March-April on y axis.
The straight line is the $y=x$ line. 
In 2019, the contribution of path disturbance has increased with time, while those related to traffic engineering have mostly remained stable (i.e. hovering around $y=x$).
The path disturbance trend is completely reversed in 2020, which also shows an increase in updates related to traffic engineering.

{\bf Takeaways.} 
The COVID-19 outbreak appears to have impacted the global routing system in several ways.
Transient routing changes have evidently decreased during the lockdown period, while traffic engineering have picked up. 
We hypothesize that several networks have frozen unnecessary maintenance and upgrades, thus resulting in a less chatty routing system. 
This gives a rare window into the composition of routing dynamics, which can help future work understanding and modeling BGP dynamics. 
\label{sec:control}

\section{Impact on data plane}

In this section, we seek to investigate whether different country-level lockdown measures impacted the data plane performance. 
Note that the form of the pandemic lockdown were different between countries -- from severe restrictions on mobility in some countries like Italy or Spain to only banning international travel in other countries like Sweden~\cite{eu_lockdown_covid19}. We use the term {\em lockdown date} to refer to the day when governments declared such measures.

\subsection{Dataset}
We use successful ping measurements collected from the RIPE Atlas platform~\cite{ripe_atlas}. For most of the countries, the pandemic lockdowns occurred during the second half of March 2020~\cite{wiki_covid19}. Hence, we choose March and April 2020 as our measurement period. Prior to our analysis, we filter our data by only considering RIPE Atlas probe pairs for which we collect more than 300 measurements per day. With this requirement, we include approximately 96\% of the (source, destination) probe pairs each day. We focus in our analysis on domestic measurements within each country. We look at countries for which we collect data both before and after the lockdown dates. 
Our filtered data covers 66 countries from across the world -- 41 from Europe, 12 from Asia Pacific region, 8 and 2 from South and North America, and 3 from Africa. 

\subsection{Impact on delay within countries}
Our goal is to understand whether we observe changes in the delay values within a country after the reported lockdown date. Hence, for each country we analyze the period {\em before} and {\em after} this date. To this end, we take the following steps. For each ping measurement, we compute the average RTT value. Next, for each country we group these values per hour and compute the median as well as the 5th and 95th percentiles. For each of these metrics, we further compute the average across all the values within one week before the lockdown date. We compute the similar averages for the first week after the same date. Finally, we compute the percentage of increase or decrease of these averages after the lockdown for each country.

\begin{figure}[h]
\centering
\includegraphics{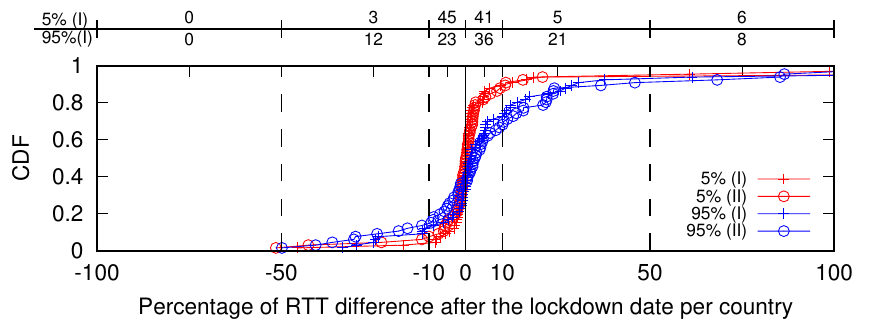}
\caption{Distribution of percentage in the RTT difference between the average values of the 5th (5\%) and 95th (95\%) after and before the lockdown date per country.}
\label{fig:cdf_procentagertt}
\end{figure}

We plot in Figure~\ref{fig:cdf_procentagertt} the distribution of these percentages. We mark with the dash lines the 10\% and 50\% increase and decrease in the RTT percentage~\footnote{Our results indicate similar trends when considering two or one week as the study period before and after the reported lockdown date.}. 
The numbers above the plot indicate the percentage of countries for which the delay has increased/decreased for the different thresholds. 
We find an increase in the 5th percentile percentage (red lines) for half of the considered countries. 86\% of these countries experience up to 10\% change in this metric.
When considering the 95th percentile (blue lines) we find that 65\% of the countries experience an increase in this percentage during the week after the lockdown date. We note, however, a high variability in this percentage. 
The increase in the 5th percentile hints at a general increase in traffic across the day (i.e. the minimum level of network utilization has increased).  

\vspace{-3mm}

\subsection{Case studies}
We further analyze selected countries for which we observe a change in both the 5th and 95th percentile delays. 
Table~\ref{tab:rtt_percentage} lists the top countries for which we observe an increase in both values. 

\begin{table}[t]
\begin{center}
\begin{tabular}{|l|cc|}
\cline{1-3}
Country (Code) & 5\% & 95\% \\
\hline
\hline
Bulgaria (BG) & 134\% & 28.82\% \\
\cline{1-3}
Kazakhstan (KZ) & 98.62\% & 0.88\% \\
\cline{1-3}
Denmark (DK) & 12.73\% & 22.55\% \\
\cline{1-3}
Italy (IT) & 5.10\% & 11.18\% \\
\cline{1-3}
Norway (NO) & 0.40\% & 37.70\% \\
\cline{1-3}
Russia (RU) & 4.76\% & 0.60\% \\
\cline{1-3}
Indonesia (ID) & 2.09\% & 5.46\% \\
\cline{1-3}
Spain (ES)  & 1.70\% & 4.78\% \\
\cline{1-3}

Germany & 1.91\% & 1.65\% \\
\cline{1-3}

\end{tabular}
\end{center}
\caption{Top countries for which the average RTT 5\% and 95\% increases after the lockdown date.}
\label{tab:rtt_percentage}
\end{table}

\begin{figure*}
\centering
    \begin{subfigure}{1\textwidth}
    \includegraphics{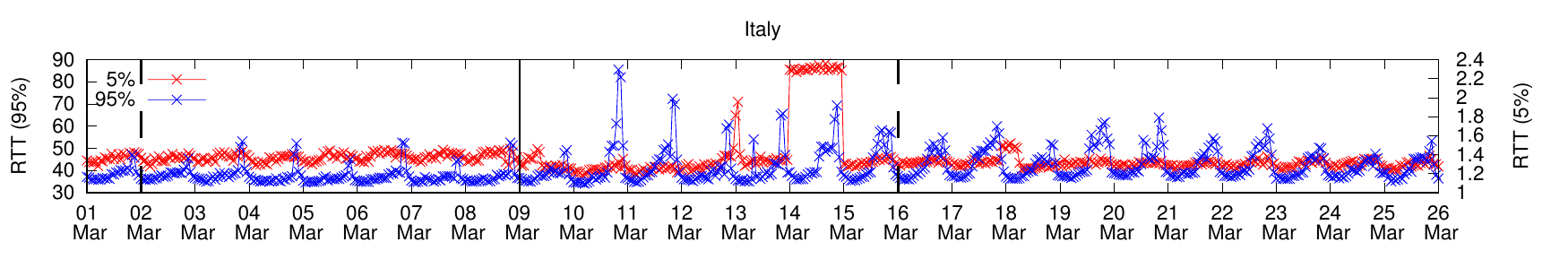}
    \label{fig:rtt_dk}
    \vspace{-5mm}
    \end{subfigure}
    \begin{subfigure}{1\textwidth}
    \includegraphics{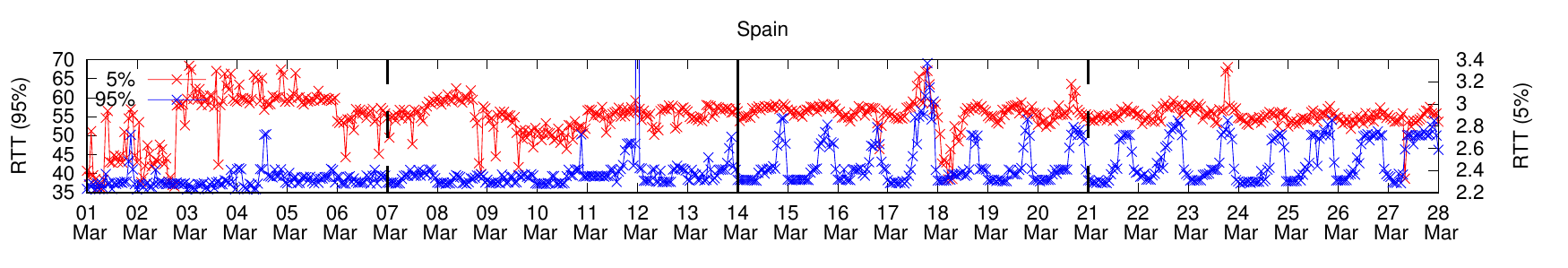}
     \label{fig:rtt_es}
     \end{subfigure}
     \vspace{-5mm}
    \begin{subfigure}{1\textwidth}
    \includegraphics{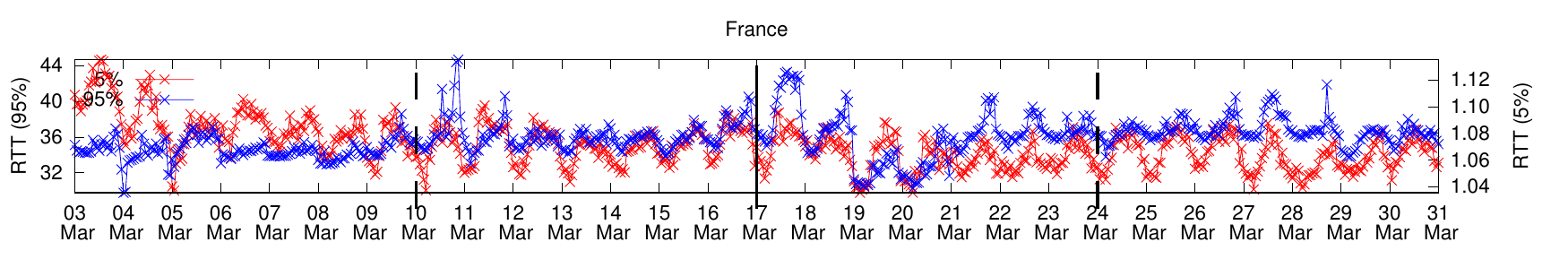}
    \label{fig:rtt_fr}
    \end{subfigure}
      \vspace{-5mm}
    \begin{subfigure}{1\textwidth}
    \includegraphics{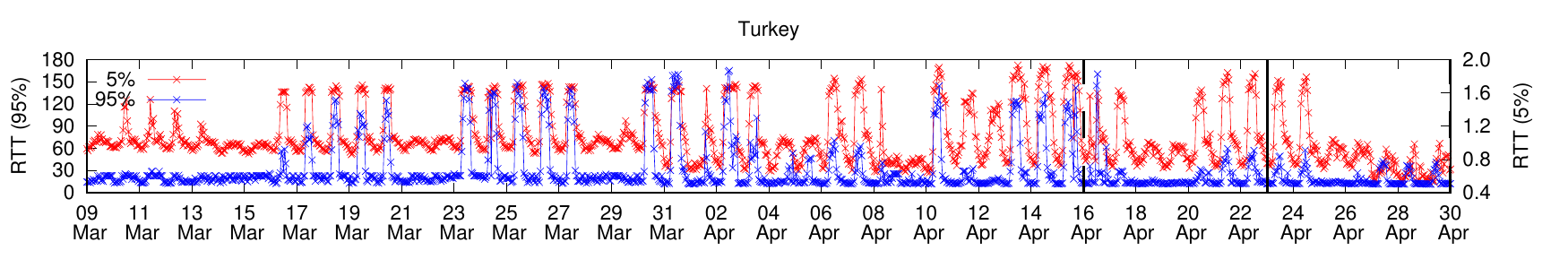}
    \label{fig:rtt_tr}
    \end{subfigure}
      \vspace{-5mm}
    \begin{subfigure}{1\textwidth}
    \includegraphics{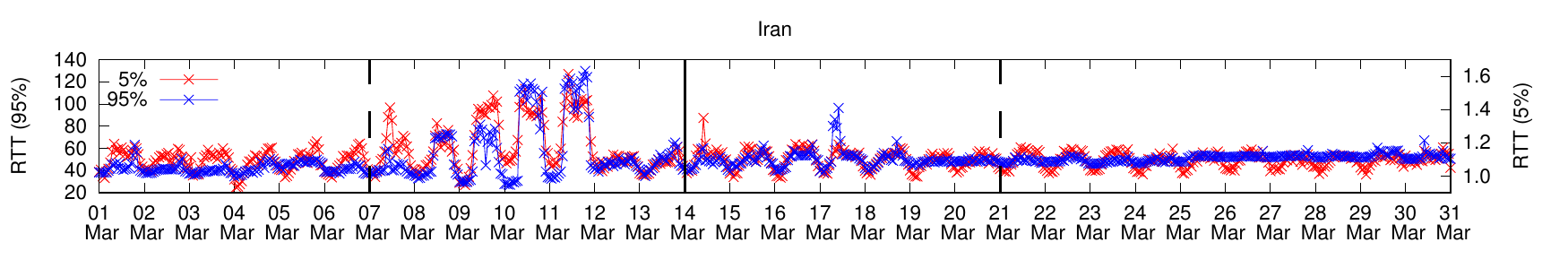}
    \label{fig:rtt_tr}
    \end{subfigure}
      \vspace{-5mm}
    \begin{subfigure}{1\textwidth}
    \includegraphics{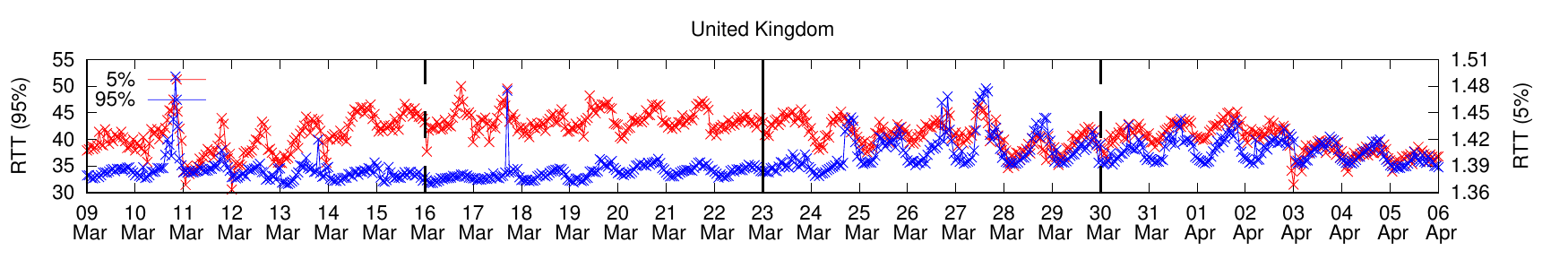}
    \label{fig:rtt_gb}
    \end{subfigure}
      \vspace{-5mm}
    \begin{subfigure}{1\textwidth}
    \includegraphics{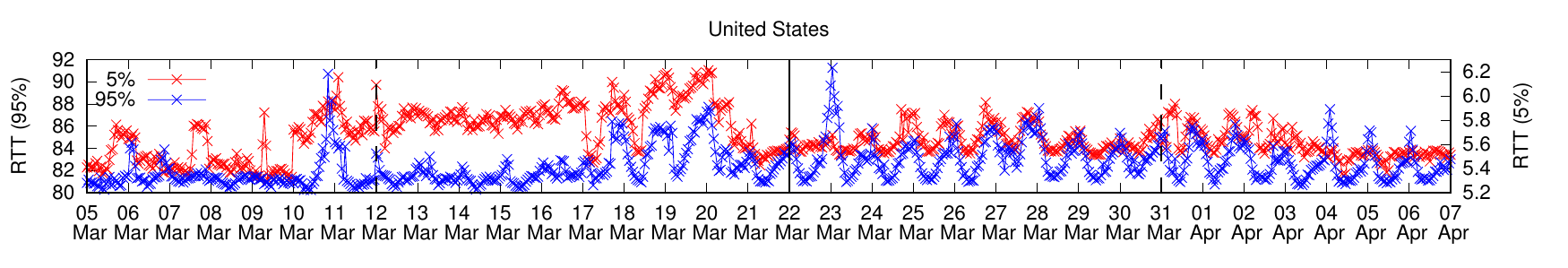}
    \label{fig:rtt_it}
    \end{subfigure}
\caption{Evolution of the 5th and 95th percentile RTT values per country.}
\label{fig:rtt_countries}
\end{figure*}

\noindent {\bf RTT values increase.} For one third of the considered countries, we observe an increase in the delay values. In Europe, Italy and Spain are two of the countries where the pandemic had severe effects. Consequently, the local governments imposed national lockdown measurements that included general confinement of the population at home, travel bans within and outside the country and closing of non-essential businesses, universities and schools~\cite{eu_lockdown_covid19}. Italy imposed these measures from the 9th of March, while Spain declared the emergency state on the 14th of March. For both countries we note a clear diurnal pattern in the 95th percentile RTT values after the lockdown dates (see the two top panels in Fig.
~\ref{fig:rtt_countries}) , which hints at periods of congestion in the networks where the RIPE probes are located.
The extent of the increase in the 95th percentile exceeds 50\% in some days.
Moreover, we note short-lived increases in the 5th percentile, i.e., during 14th and 17th of March for Italy and Spain, respectively. 

\noindent {\bf RTT values decrease.} Our analysis shows that for some countries, the RTT values actually decreased after the lockdown date. We manually inspect the top countries, and find that for most of these the decrease is due to level shifts in the RTT that could be caused by path changes or loss of high-delay monitors (i.e. measurements artefacts). 
In some cases the decrease stems from the fact governments in some countries have taken a gradual approach to lockdown attempting several policies on the way.
Thus rendering the final lockdown date inconsequential to the impact on the Internet.
Turkey is an example of a such country 
We find that delay has started to increase \emph{prior} to the reported lockdown date. 
As shown in Fig.~\ref{fig:rtt_countries} the 95th percentile of the RTT values increased up to four-folds from the second week of March. 
We also observe diurnal patterns during weekdays in both the 5th and 95th percentiles. 
A closer analysis of the pandemic response timeline shows that the Turkish government have imposed gradual measures that spread across more than one month. Schools and universities were closed for three weeks from the 12th of March, entry to some provinces was banned on the 3rd of April, and a four-day lockdown period was announced from the 23rd of April~\cite{tr_timeline_covid19}. Our collected data clearly indicates the impact of this period on the delay values. 
However, considering only the four-day lockdown period indicates a decreasing trend in delay.

Iran was one of the first countries, after China, to report a high number of infections.
The government imposed a national lockdown on the 14th of March.  
The inter-Iran delay timeseries (Figure~\ref{fig:rtt_countries}) shows a short-lived increases in the RTT values six days prior to the lockdown dates. A closer analysis shows this increase is linked to one of the RIPE probes located in AS60256. Comparing the first week of March with period after the lockdown day, we do not record significant differences in delay.

For countries like France and Austria, we also find a slight decrease in the delay values immediately after the lockdown date.  Similar to Italy and Spain, France was one of the countries with a high number of reported infections and a national lockdown from the 17th of March. The collected data shows a fluctuation in the RTT values following this date for both the 5th and 95th percentile. However, we note that these values remain at approximately the same levels throughout March. In the case of Austria, we also record a small drop in the RTT values; the 95th and 5th percentile drop with 1.8\% and 0.1\%, respectively. 

The Swedish pandemic response was completely different than most European countries. Sweden did not impose a lockdown and just issued a series of recommendations as a response. Our data shows a low impact on the delay within Sweden.
The delay has remained stable throughout both March and April. 

Argentina imposed lockdown measures similar to the ones imposed in Italy and Spain on the 19th of March. Initially planned until the end of March, this period was later extended to the third week of April~\cite{argentina_covid19}. 
We observe as with other countries diurnal pattern in the 95th percentile following the lockdown date. Moreover, during the last days of March and first half of April we notice a significant increase in delay. We can thus conclude that the pandemic outbreak had an impact on the delay within Argentina. 

 
In the case of Ireland, United Kingdom and United States we find an increase in the 95th percentile and decrease in the 5th percentile values. In Ireland, the government initiated the lockdown period from the 12th of March along with measures like closing universities and schools and shutting down non-essential businesses from the 24th of March. Our results indicate an increase of 30\% in the 95th percentile in the week following the lockdown date. 
Moreover, our analysis shows signs of congestion during the same period and the following weeks.
United Kingdom and United States imposed gradual measures to counter the outbreak. 
In the UK, a stay-at-home recommendation was given as early as the 16th of March, while a stricter requirement was issued on the 23rd of March~\cite{gb_timeline_covid19}. 
The response in US has varied across states. 
States like California, Massachusetts, and New York imposed a stay-at-home order from 19th and 20th of March.  
A similar restriction was imposed already on the 17th of March in the  San Francisco Bay Area~\cite{us_sfo_covid19}. However, states like Iowa or Wyoming did not impose such restrictions. For both the United Kingdom and United States, we note during the measurement period an increase in the 95th percentile RTT values. At the same time, we observe signs of congestion in the probe networks for both countries from around the lockdown date. For both these countries we also observe an increase in the 5th percentile   during March on several days.

{\bf Takeaways.} 
Our analysis of the RTT values as seen from the RIPE Atlas measurement platform indicates that pandemic outbreak appears to have an impact on the intra-country delay. While this effect varies between countries and may be influenced by other underlying factors, we find clear indications of periods of congestion following the lockdown dates across multiple countries.
Hikes in delay varies across countries which can be attributed to differences in the underlying infrastructure and the fraction of telecommuting population.

\label{sec:Data}

\section{Outages during the outbreak}
In this section, we employ two different approaches to investigate whether the lockdown has led to an increase in Internet outages.
The first is based on passive measurements from RIPE Atlas~\cite{ripe_atlas}, a global measurement infrastructure of over 10,000 probes that is operated by RIPE.  
The second analyzes operators' reports in the outages mailing list.~\footnote{https://puck.nether.net/mailman/listinfo/outages} 

\subsection{RIPE Atlas outages}
We leverage DISCO, a tool that monitors the stability of RIPE Atlas probes, to infer outages~\cite{shah2017disco}.
Every RIPE Atlas probe maintains a reverse SSH connection to a set of servers called controllers in the RIPE infrastructure. 
These sessions are used for sending commands to the probes. 
If the controller does not receive a keep alive from a probe for a minute, it tears down the SSH connection and marks the probe as unresponsive.
DISCO identifies and correlates such events to infer outages in an AS or a geography.

Comparing the outages in the first four months of 2020 to those in 2019 shows an evident increase.
1227 prefixes from 110 ASes had an outage involving 1809  unique  probes in 2020 compared to 61 prefixes from 31 ASes (105 unique probes) in 2019.
Most of the increase took place during March and April.
On average outages lasted shorter in 2020 than in 2019 - about 40 minutes compared to over three hours.   


\begin{figure}[t]
\centering
\includegraphics[width =1.1\columnwidth, ]{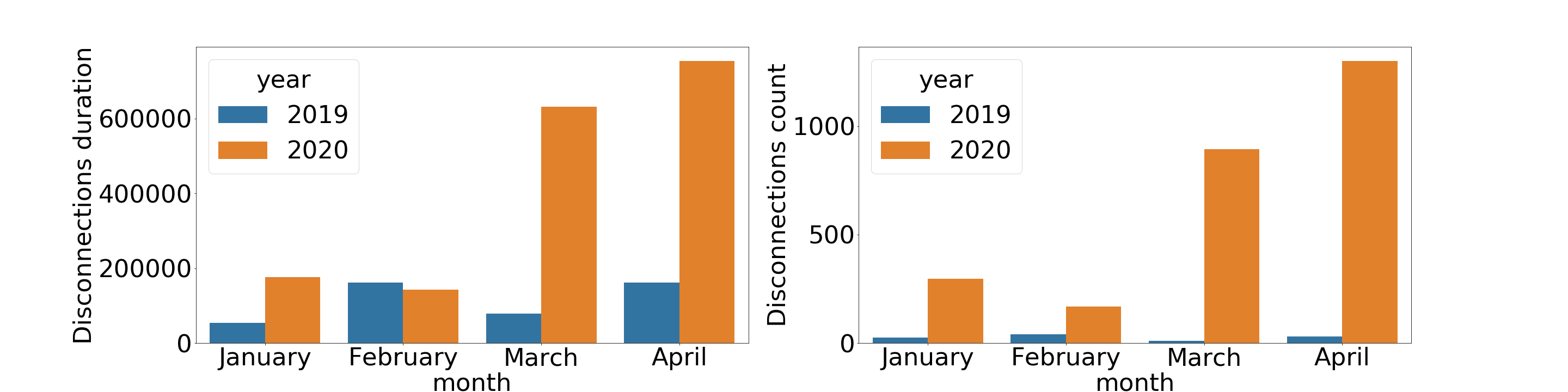}
\caption{Aggregate outage duration (left) and outages count (right) both after filtering.}
\label{fig:outages-filtered}
\end{figure}

While we do not know why outages have generally increased in 2020, we suspect that it could be related to a few ASes that had a disproportionately high number of outages.
To control for this, we filter the top 25\% of prefixes in terms of either disconnection duration or number of outages.
Figure~\ref{fig:outages-filtered} shows the number of outages and their duration after filtering.
The increase in the number and severity of outages in March and April 2020 is clearly visible even after this filtering.
The average outage duration drops to ten minutes for 2020 and just over an hour for 2019 after filtering.

The affected prefixes come from 35 countries from all continents.   
Iran dominates outages across the four months accounting for 29\% and 19\% of the outages in the first two months and the last two months, respectively.
This is evident since early January so it is not clear whether it is related to the outbreak.

European countries are over-represented in March and April.
The fraction of outages that involves a European country has risen from 38\% in January-February to 51\% in March-April.
In the first two months of 2020,  46\% of outages in Europe took place during typical maintenance hours. 
This fraction dropped to 40\% in March-April. 
Hence, the increase in outages during the outbreak can be blamed on a marked increase in both non-maintenance related as well as maintenance related faults.

Although a detailed root cause analysis of the inferred outages is outside the scope of this work, we manually investigate some of the outages in Europe that took place outside typical maintenance hours. 
We particularly flag two outages that affected home users and lasted for a relatively long period and thus may have had impact on people working from home.
The first affected Virgin Media (AS 5089) in the afternoon of April 27th and lasted for about 6 hours.
During these hours RIPE probes had intermittent connectivity, experiencing a total of 7 disconnects each lasting between 10 and 15 minutes. 
At its height, 94 of the 97 RIPE probes in the affected AS were disconnected.
This outage was reported by various media outlets due to its significant impact and was attributed to a fault in Virgin Media's core network.~\footnote{https://www.bbc.com/news/technology-52448607} 
The second outage affected an Austrian cable provider, Kabelplus, in the morning of April 1st and lasted for about 20 minutes.
During this outage 7 of 12 probes in Kabelplus's network, which come from two different ASes 8339 and 8559, were disconnected.

\subsection{Outages mailing list}
We browsed the reports that were posted to the outages mailing list in the months of January to April in both 2019 and 2020.
We then singled out all reports that are related to Internet outages (i.e. we left out voice), which were corroborated by postings from several individuals or were confirmed by the affected service providers. 
Note that the outages mailing list is often used for reporting planned maintenance and significant outages and mostly reflect the view of operators from North America.

\begin{table}[h!]
\centering
\begin{tabular}{ |c|c|c|c|c| } 
\hline
{\bf year} & {\bf January} & {\bf February} & {\bf March} & {\bf April} \\
\hline
2019 & 8 & 5 & 7 & 4
\\
\hline
2020 & 6 & 4 & 16 & 11
\\
\hline
\end{tabular}
\caption{The number of outage reports}
\label{tab:outage}
\end{table}

Table~\ref{tab:outage} shows the number of outages that were reported in different months. 
Although the numbers are small, we still observe a clear increase in the number of outages reported in March and April 2020. 

Besides the increase in outages, what is also interesting is the type of outages.
The second half of March involved a few reports of increased packet loss and delays between several US providers. 
Such failures were probably caused by congestion or lack of spare capacity following a failure of a primary conduit.  
Youtube and Google services degraded for about half an hour on March 26th.
Google later published a blog post indicating that its backbone network is designed for absorbing the increase in traffic due to the covid-19 crisis.~\footnote{https://www.blog.google/inside-google/infrastructure/keeping-our-network-infrastructure-strong-amid-covid-19/} 
The reporting of increases in packet loss and delay continued in April.
There were also reports related to collaborative and video conferencing platforms. 
On April 21st, several major tier-1 providers experienced packet loss in the US West Coast.
These include AT\&T, Level-3/Century Link, Hurricane Electric, Telia, Cox and Comcast.
The root cause turned out to be the failure of multiple fiber links, which was exacerbated by a lack of spare capacity due to the increasing traffic demand. 
This type of failure is concerning, since it illustrates the potential risk for cascading failures as traffic demand increases.  

{\bf Takeaways.} 
The number of network outages has increased during the lockdown period. 
The root causes include maintenance, congestion and lack of network capacity upon the loss of a primary link.
Although most outages were short, some outages lasted hours and had likely impacted remote workers.  
\label{sec:outages}

\section{Impact on DNS}

Having investigated the performance of control and data planes and the stability of the Internet during the lockdown period, we move next to examine the impact on applications.
In this section, we compare the performance of accessing DNS root servers.

\subsection{Dataset}

We leverage data collected by the RIPE NCC DNS monitoring service (DNSMON)\cite{amin2015visualization}.
DNSMON measures the performance of DNS from a set of RIPE Atlas anchors to all root DNS servers since 2001. 
The set of anchors coverage is geographically large as possible and each anchor runs the same set of measurements towards the same targets and periodically reports the results back to the RIPE Atlas infrastructure.
This involves both DNS lookup and traceroutes to all root servers over both IPv4 and IPv6 and using both TCP and UDP.
We focus on data that covers the two periods (January-April) 2019 and 2020.  



\subsection{DNS performance}

\begin{figure}[t]
\vspace{1mm}
\centering
\includegraphics[width =\columnwidth]{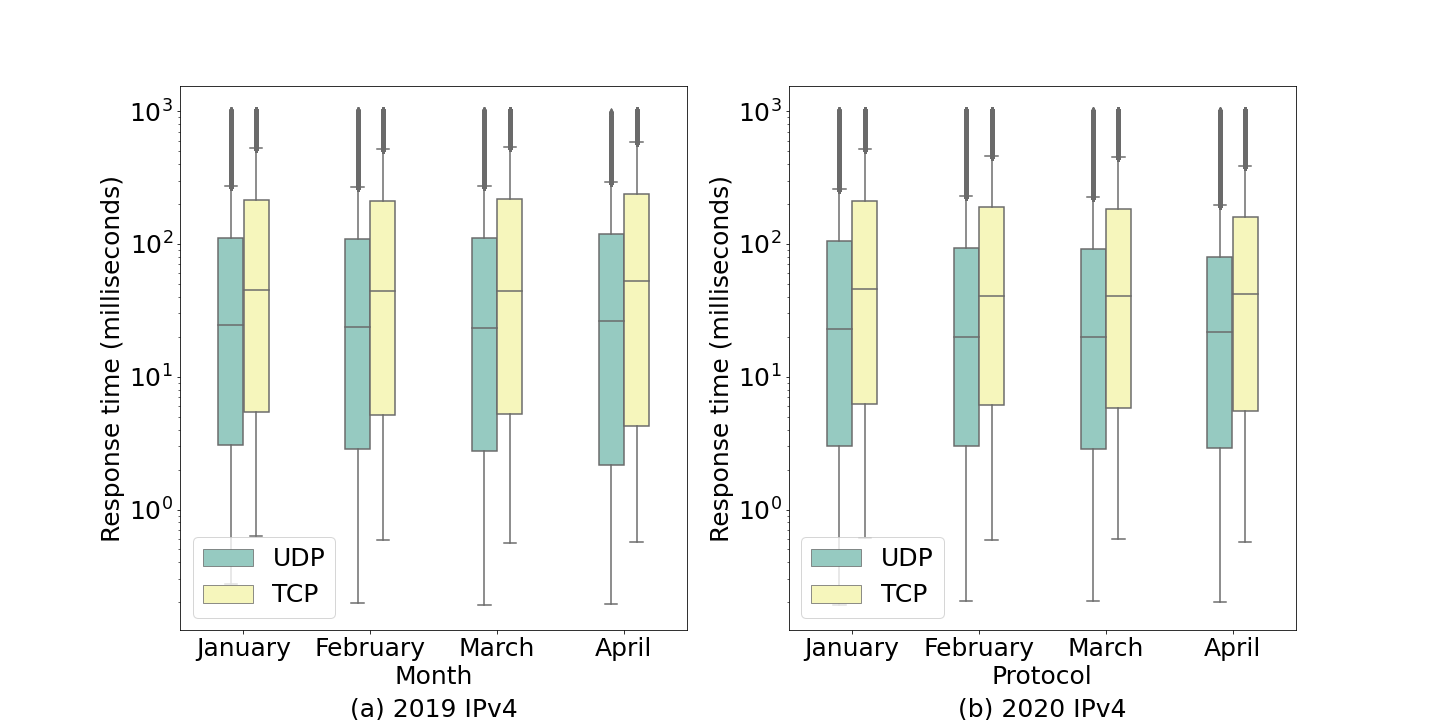}
\caption{Root DNS response time over IPv4 from January to April 2019 and 2020.}
\label{fig:dns_response}
\vspace{2mm}
\end{figure}

Figure~\ref{fig:dns_response} shows that the root DNS response time over IPv4 has been relatively stable since 2019.
we record a slight increase, about a few milliseconds, in the median response time in April 2020.
The shape of the distribution, however remains similar to the previous months. 
Lookups over IPv6 are also stable albeit faster with a median response time below 20 milliseconds. 
Moura et al. have reported similar results when measuring the performance of .nl TLD~\cite{moura2020coronavirus}. 




\begin{figure}[h]
\vspace{1mm}
\centering
\includegraphics[width =\columnwidth]{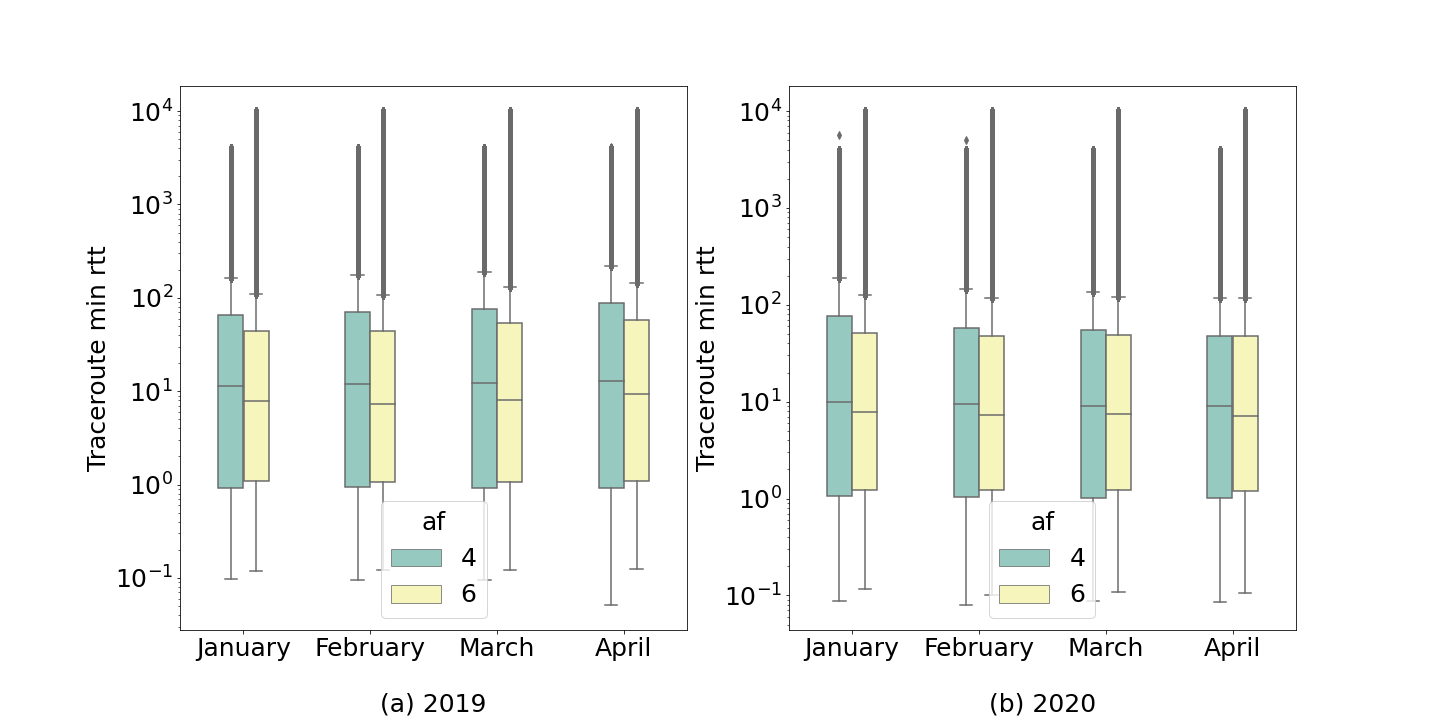}
\caption{Root DNS traceroute minimum rtt from January to April 2019 and 2020.}
\label{fig:dns_traceroute}
\vspace{2mm}
\end{figure}

Similarly Figure~\ref{fig:dns_traceroute} shows no changes in traceroute minimum RTT to Root DNS server during the lockdown period.
Interestingly, the 75th percentile IPv4 RTT has decreased in April to match that of IPv6. 
The consistent minimum traceroute RTTs confirm the DNS lookup results above. 

The performance stability of Root DNS servers is a proof of the resiliency and scalability of this critical component of the Internet.
The use of anycast combined with caching on resolvers side means that traffic to Root DNS servers is relatively lower than traffic to popular open resolvers (like Google Public DNS, Quad9, Cloudfare, etc.). Therefore, it could be also interesting to take a look at open resolver performance during the lockdown.  

{\bf Takeaways.} 
The performance of DNS root servers has been consistent during the lookdown period regardless of the IP protocol version or the transport protocol in use.   
\label{sec:dns}

\section{Case study: video conferencing }
Continuing on evaluating the impact on application, we study next how the lockdown has impacted a European video conference network provider with global presence.
The video network customers fall in two categories, 1) enterprise video conferencing customers and video conferencing service providers hosting their video infrastructure and 2) subscribers to the recording and streaming service used by enterprise customers, education, and health institutions. Both applications are covered in the recorded data.

\subsection{Dataset}
We collect our measurments from a global video conference network provider operating a QoS managed network built of dedicated links between 11 PoPs in Europe, North  America and Asia Pacific (APAC). 
The network provides QoS managed last mile connections from their PoPs via approximately 25 commercial IP transit providers and presence at 10 Internet Exchange points with more than 1500 BGP peering connections. 
Each PoP hosts network equipment and compute capacity providing a private cloud used by their customers to host video conferencing technology from Pexip, Cisco and other video infrastructure providers. 
We collect flow data using Netflow from the network operator's PoPs in Frankfurt, Amsterdam, Oslo, London in Europe, Asburn, San Jose and Toronto in North America, and Sydney, Tokyo, Hong Kong and Singapore in APAC.   
We focus on two measurement periods.
Before the lockdown, from the 1st to the 13th of February.
After the lockdown, from the 15th of March to the 30th of April. 
Besides, the Netflow dataset, we also analyze the underlying data that the video provider is using for accounting purposes.
This includes traffic volumes per customer that is aggregated every 30 minutes and spans the period from November 2019 to May 2020. 
The Netflow measurements give an insight into traffic flows at the IP level, while the billing data gives an idea about traffic volumes per customer.  


\subsection{Impact on traffic volumes}
\begin{figure}[t]
\centering
\includegraphics[width =1.0\columnwidth, ]{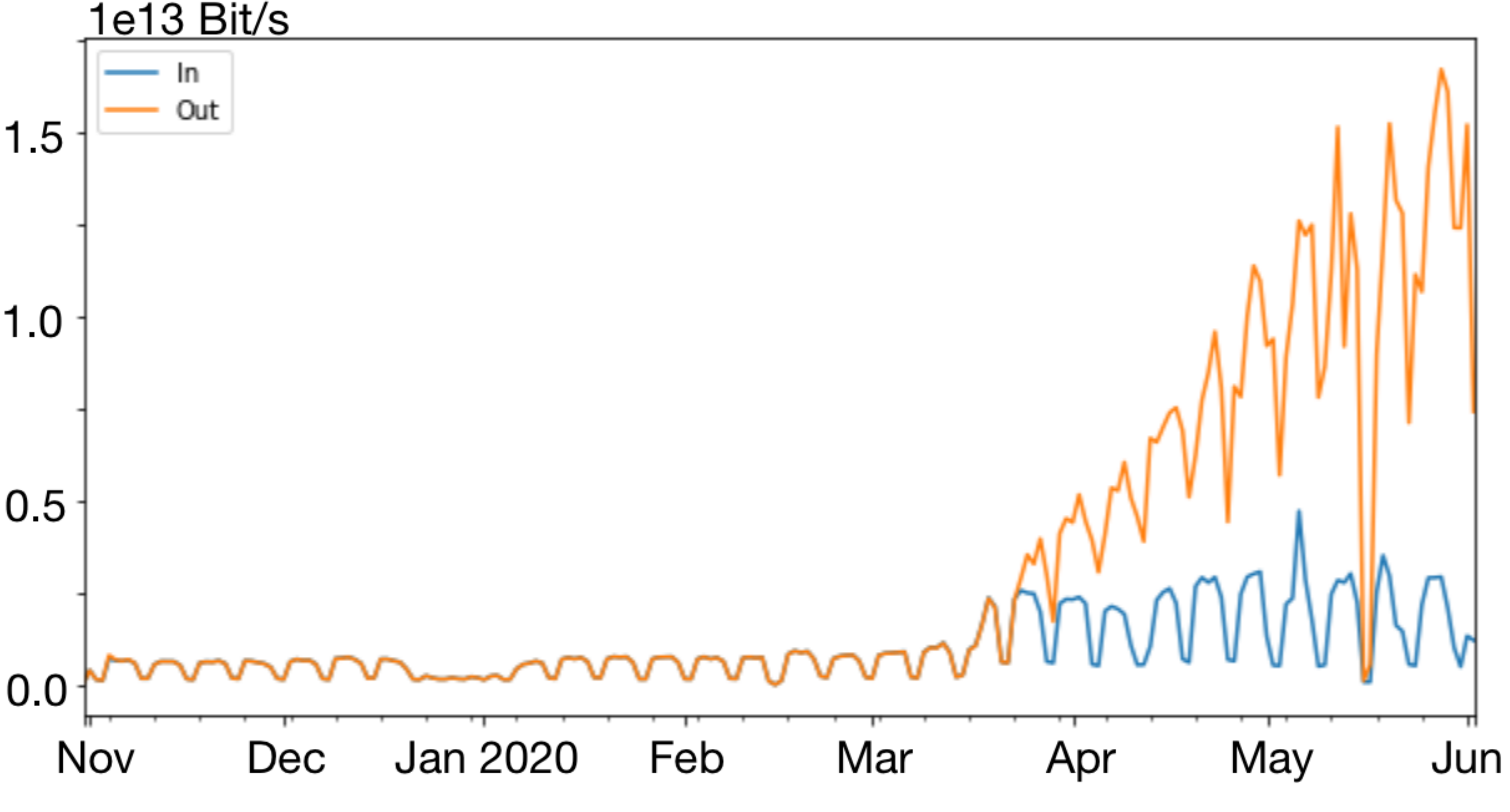}
\caption{Video conferencing traffic growth in a global video network.}
\label{fig:mns-timeseries}
\end{figure}

Working from home and attending school during the pandemic have significantly affected usage of all types of video conferencing users. This can be seen in the unusual growth in video conferencing traffic of the network provider in Figure ~\ref{fig:mns-timeseries}, increasing the amount of combined traffic for all interfaces from a steady average of 1 Tbit/s input and output traffic per day before the pandemic to almost 20 Tbit/s during the pandemic with a rapid increase from mid-March when the lock-down started in several countries where the service provider operates. Note that this dataset count traffic both on customer interfaces and on transit/internet exchange interfaces, so actual transported traffic grows from 0,5 Tbit/s to approximately 10 Tbit/s. The traffic continues to grow during the pandemic. 

Another observation is that the amount of outbound traffic is growing faster during the pandemic. The trend was visible before the pandemic as video conferencing upstream video often use lower resolution and users sometimes turn off their own video transmission when not talking, while downstream video have higher quality and is usually on during the entire conference. During the pandemic, this trend is stronger than we can explain with more people attending video conferences. We found that the conferencing service is increasingly combined with streaming sessions, where large groups of users attend a single live or stored stream, resulting in an accelerated growth in outbound traffic. Note that the general video conferencing traffic also shows strong growth.

\subsection{Sources of growth}

\begin{figure}[t]
\centering
\includegraphics[width =1.0\columnwidth, ]{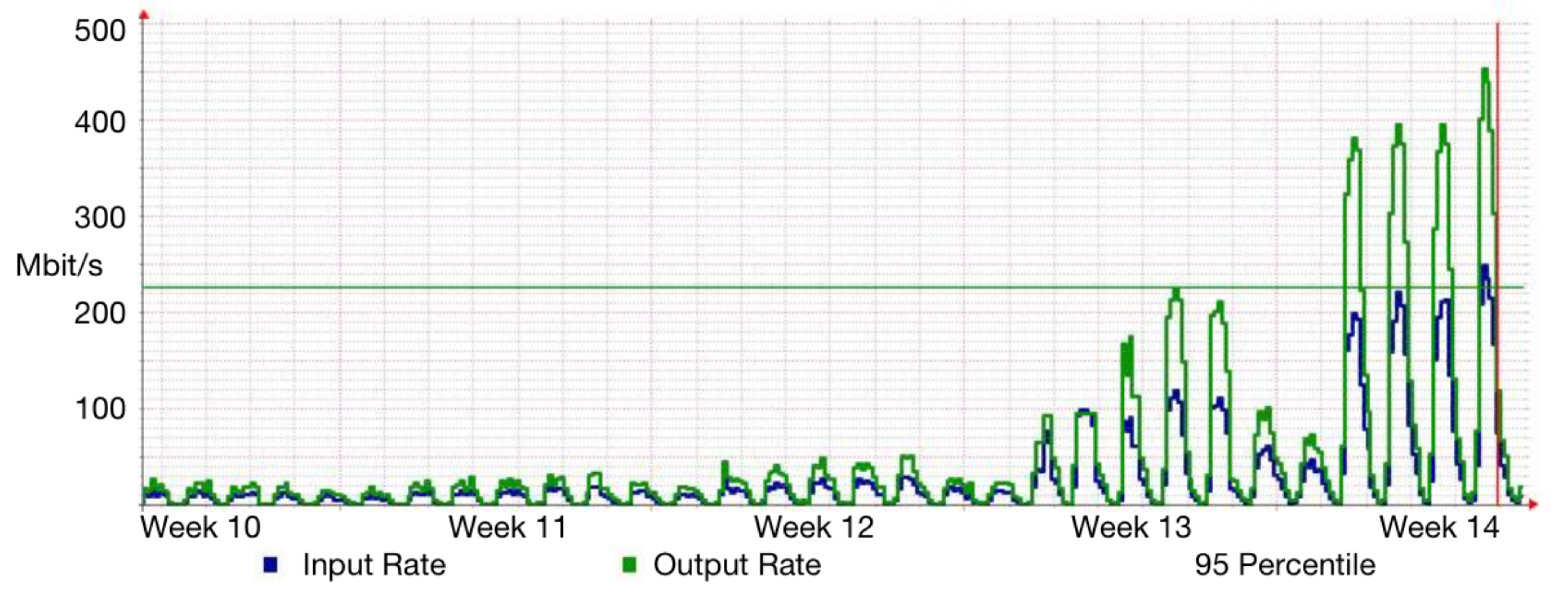}
\caption{Video conferencing traffic growth for a telehealth provider in the US.}
\label{fig:video-application}
\end{figure}

We have identified two sources of the increased traffic: streaming and video conferencing. Streaming has the highest impact, most notably for outbound traffic. Video recording hours, which is used before creating a video stream, had an increase of 208\% from February to March 2020, and streaming use increased 1100\%. 
Figure ~\ref{fig:video-application} shows a dramatic example of traffic growth for an American telehealth customer in the video network providing remote medical consultations, going from an average of 25 Mbit/s (95 percentile) regular traffic from March 1 - March 13, 2020, increasing to 50 Mbit/s from March 14 - March 20, 225 Mbit/s from March 21-26 and exceeding 450 Mbit/s in April 2020. Sustaining a growth of up to 18x in the course of a few weeks represents extraordinary demands on network and compute facilities and the service provider and suppliers had to work double shifts to upgrade links, configure, ship and install equipment to meet the demand. Similar growth has been reported by other video infrastructure vendors \cite{starleaf}. 

A detailed study of the growth in flows compared to total traffic shows that early during the pandemic, most video service providers in the network reduced video resolution from high definition (HD) to standard definition (SD). This method reduced the growth in traffic but the number of flows increased with the growing demand. For example, we find that a video infrastructure customer in the dataset experienced 400\% growth in the number of video flows from February to March, while traffic grew only 50\% in the same period which ensured the service provider had sufficient network and compute capacity to cope with the demand. We confirmed the reduction in resolution with the customer. 
The provider has also confirmed that many video service providers did reduce video conferencing resolution, which indicates that the real traffic growth could have been even higher. 


{\bf Takeaways.} 
Our analysis shows that video conferencing traffic is growing rapidly during the pandemic, and the use of streaming services dominate the outbound traffic which can be explained with the use of the infrastructure for education (several universities use the streaming platform), one-to-many conferences and streaming events during lockdown. Reduction from HD to SD resolution by many providers has slowed down the traffic growth, but increased volume can still be seen from the number of netflows per customer. The highest traffic growth for an individual customer is 18x during the pandemic compared to before the pandemic. 
\label{sec:video}

\section{Case study: mobile network data traffic}

We leverage a country-wide deployment of stationary probes in a western country to measure  mobile data uplink  and  downlink speeds. 
Each  probe  is  a  single-board  computer  that  connects to one or more mobile networks, via miniPCI modems that support up to LTE CAT-6, using commercial subscriptions. To measure speed, we use a command line client for testing Internet speed using Ookla’s speedtest.net. The test is essentially based on downloading and uploading files of an  increasing  size  from and to  Ookla’s  servers and used that to estimate available bandwidth.  
The test runs three times a day at 2:00 am, 2:00 pm and 7:00 pm local time to capture different traffic profiles.   
The probes connect via different frequency bands, depending on available coverage.  
These include 800 MHz, 900MHz,  1800  MHz,  2100  MHz  and  2600  MHz.   

We collect speed measurements for two different mobile networks Op$_1$ and Op$_2$ from 102 probes in the months of February and April 2020. These two periods represent the situation before and during pandemic-related restrictions in the country. 
The 102 measurement nodes are located in 40 different municipalities.

\begin{figure}[h]
\centering
\includegraphics[width = 1\columnwidth]{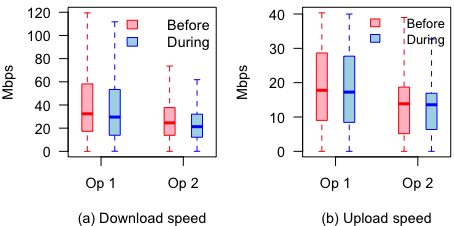}
\caption{Drop in speed with increased load on the networks.}
\label{fig:download_upload}
\end{figure}

\begin{figure}[h]
\centering
\includegraphics[width = 1\columnwidth]{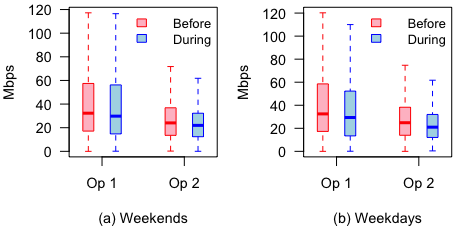}
\caption{Download speed during weekends and weekdays.}
\label{fig:weekend_days}
\end{figure}

Both operators have reported an increase in mobile data traffic of up to 25\% following the restrictions. ~\autoref{fig:download_upload} shows that with the increased load, an overall drop of 8.7\% and 13.4\% occurred in the download speeds for Op$_1$ and Op$_2$, respectively in the \textit{during} period. The drop in upload speed is more limited, with 2.9\% and 2.1\% for the two operators. 
Moreover, we observed comparatively more drop on weekdays.~\autoref{fig:weekend_days} shows that on weekend the median drop in speed for Op$_1$ is 7.7\% and for Op$_2$ it is 8.7\%. However for weekdays the observed median download speed goes down by 9.5\% and 15.57\% for the two operators, respectively.

\begin{figure}[h]
\centering
\includegraphics[width = 1\columnwidth]{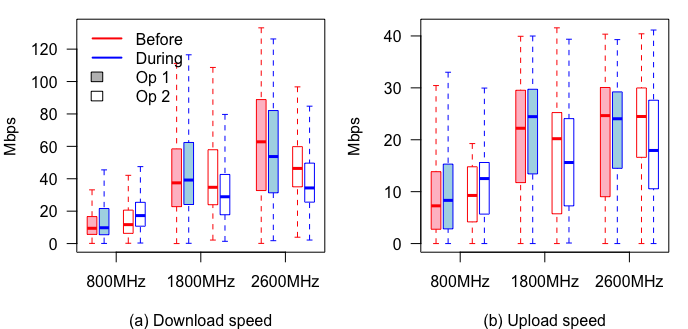}
\caption{Download and upload speeds before and during the lockdown period with probes operating at different frequencies.}
\label{fig:freq_download_upload}
\end{figure}

~\autoref{fig:freq_download_upload} shows the change in download and upload speed split on the frequency bands that are mostly used for LTE traffic by Op$_1$ and Op$_2$.
The figure shows how both download and upload speeds are affected differently in different frequency bands. 
Both operators experience an increase in upload and download speeds for connections in the 800 MHz band. 
For the 1800 MHz band the results are mixed, while both operators experience a significant decrease in speeds in the 2600 MHz band.


Mobile operators generally use higher LTE bands (2100 MHz and 2600 MHz) to provide coverage in densely populated areas where people live and work. 
Lower bands (sub-GHz) have better propagation characteristics and are typically used to provide  coverage along roads and in more remote areas.
We believe the results in \autoref{fig:freq_download_upload} can be directly explained by the changes in movement patterns resulting from the pandemic-related restrictions, with people generally staying at home and working from home offices.
These lead to increased mobile data traffic in areas where people live, and less traffic where people move.



\begin{figure}[h]
\centering
\includegraphics[width = 
1\columnwidth]{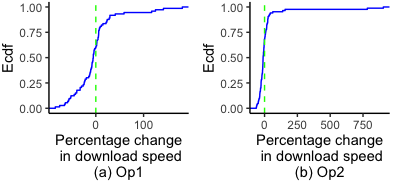}
\caption{Change in the download speed, observed, in the \textit{during} period at different probes.}
\label{fig:nodes}
\end{figure}

To further understand reasons behind the observed variations in speed, we look at results on the granularity of probes locations. 
~\autoref{fig:nodes} shows the change in download speed for each connection, when comparing the periods before and during lockdown.
We observe that 59\% of connections in Op$_1$ and 66\% of connections in Op$_2$ experience a reduction in download speeds.
The reduction is 50\% or higher for 18\% and 15\% of the total connections, of Op$_1$ and Op$_2$ respectively.
The remaining connections see an increase in download speeds, which are more than 100\% for some connections.
 
\begin{figure}[h]
\centering
\includegraphics[width = 1\columnwidth]{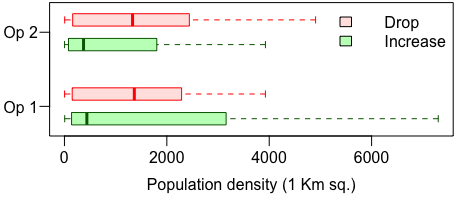}
\caption{Population density at locations with increase and drop in the download speed.}
\label{fig:pop_density}
\end{figure}

\autoref{fig:pop_density} shows how the changes in download speeds correlate with the population density of the area where the measurement is taken.
Measurement probes that see an increase in download speeds are mostly situated in areas with lower population density, while those that see a drop in speed are located in more densely populated areas. 
The median population density in 1 $km^2$ squares where probes see increased download speeds is 462 and 414, while it is 1367 and 1363 in areas where download speeds drop, for Op$_1$ and Op$_2$ respectively.
These observations show how the mobile networks struggle to cope with the increased traffic in residential areas during the lockdown, while the changes in movement patterns give plenty of capacity in other areas.

{\bf Takeaways.} 
The lockdown has an evident impact on the performance of the measured mobile networks.
In the worst case, some probes achieved only half of the speed that was achievable before the lockdown.


\label{sec:speed}

\section{Discussion}
We have provided a multi-prospective account of  Internet performance during the first lockdown period. 
In the following, we discuss our findings, their implications and. limitations   

\noindent{\bf The Internet during the lockdown.} 
The impact of the lockdown period on the Internet is visible on various performance metrics.
Working and schooling from home have lead to a dramatic increase in video conferencing. 
Performance aspects of access network, both fixed and mobile, has noticeably degraded during the lockdown.
While we have not analyzed the general mix of traffic during the epidemic, the case study of the global video providers shows a significant growth in both video streaming and conferencing services around the world.

The data plane measurements in Sec~\ref{sec:Data} show that , in countries across the world,  both low and high percentile delays have increased in the lockdown months.
This indicates the presence of persistent congestion and higher peak traffic volumes.
The reduction in download speeds over mobile networks hints that mobile networks are also running close to their capacity during the lockdown. 
We also measure performance improvements in areas with less population as well as over frequencies that are deployed to support daily commuters. 
Besides the degraded data plane performance, outages have also increased.

Despite performance degradation due to the massive increase in traffic volumes, the core of the Internet architecture including BGP routing and DNS have been fairly stable. 
In fact, BGP dynamics have decreased during the lockdown.

Overall, the Internet seems to have managed to absorb changes in traffic pattern and volumes reasonably well.
While a more detailed analysis is needed in order to fully understand this, we see signs of clever network management, adaptive traffic strategies and abundance of spare capacity.

\noindent{\bf Opportunities.} 
The lockdown period has pushed the Internet to a new operating point, which can help answering several questions about the Internet architecture. 
Differences in data plane performance across countries could give a window into evaluating and comparing the state of Internet infrastructure in different parts of the world.
The outages that took place during the pandemic deserve further attention, since they can help revealing structural weaknesses and vulnerabilities.
For example, the US west coast outage that is briefly described in Sec~\ref{sec:outages} hints at presence of dependencies between operators that are only visible at times of extreme overload.
The decrease in BGP dynamics can help improving our understanding of the composition of BGP activity which can lead to better anomaly detection algorithms and ultimately a better interdomain routing.
Finally, the Internet performance measurements can be used as a proxy to inform efforts that aim to understand the impact of lockdown on work and education in different countries. 

\noindent{\bf Limitations and open questions.}
This paper has presented several key findings about the first lockdown period, but has also left many questions open.
These questions warrant a further investigation.
An interesting aspect of the data plane analysis is performance of inter-country routes.
This is particularly interesting in markets with tight integration like Europe. 
We have chosen to focus on intra-country latency because we believe that most home office and schooling traffic remain within each country borders.
An in-depth analysis of outages, using active and passive approaches (e.g. ~\cite{quan2013trinocular,richter2018advancing}), could shed more light on root causes and dependencies between network providers.

\label{sec:discussion}

\section{Related Work}
The academic and operators community has been following closely the possible effect of the pandemic on the Internet performance. We further summarize and contrast with our work the main reports related to the pandemic.

Several studies that characterize the data plane have been published in the last few months. Notably, Candela {\em et al.}~\cite{candela} used data collected through the RIPE Atlas measurement platform to study the impact of the pandemic on latency and packet loss for Italy and report significant increase in both metrics.
In the COVID-19 delay-based short study~\cite{ripe-delay-corona} published on RIPE Labs, the authors rely on data collected from the same platform to study  change in RTT from major ISPs in Europe to the Google network (AS15169), two large IXPs (AMS-IX and DE-CIX), and the other network(s) 
these major networks are most dependent on according to the AS hegemony. The study reports congestion signs as seen in RTT values collected from probes in the Telecom Italia network towards AMS-IX. We also rely on RIPE data for assessing the data plane performance. Our analysis, however, covers countries that from all continents, and focuses on intra-country delay. We find evidence of congestion periods in multiple countries during the lockdown periods. 
Feldmann et al.~\cite{feldmann2020lockdown} investigated, using data from a couple ISPs and three IXPs, changes in traffic volumes and patterns during the first wave of the pandemic. 
They measured a 15-20\% increase in traffic and identified changes in traffic pattern, but also concluded that the Internet has largely coped with the lockdown.
A paper from Facebook has pointed to similar conclusions from the prospective of Facebook's edge network~\cite{bottger2020internet}.
It also to variations in performance showing that the less developed parts of the world suffered more performance degradation.
Liu et al.~\cite{liu2020characterizing} looked into the impact of the lockdown period on the Internet performance in the US. 
Their results highlighted increasing in traffic volumes, heterogeneous performance and the ISPs rush to upgrade and augment their connectivity.     
Using data from a mobile operator in the UK, Lutu et al.~\cite{lutu2020characterization} investigated changes in mobility and traffic patterns. 
They noted a drop in data traffic and an uptick in voice traffic.
Along the same lines, Rajiullah et al. used end to end measurements from the MONROE platform to look at the performance of nine mobile operators in Europe   ~\cite{rajiullah2020mobile}.
Despite a short term degradation in performance, these operators seem to have coped well with the lockdown traffic. 

The video conferencing industry has reported high traffic growth. StarLeaf provided key observations before and during the pandemic in their trend report~\cite{starleaf} that comprises traffic report for their operation in France, Germany, Italy, Sweden, United Kingdom and United States. 
The report shows that sharp increases in video usage for all nations can be traced to the date of lockdown, or instructions to start working from home. 
Cloudflare~\cite{cloudflare2} reported an 1.5x performance increase in Internet traffic. Another study \cite{cloudflare2} observed that when people pause in their daily activities, the Internet goes quiet and show this with examples from the event \#ClapForNHS in UK and Ramadan in the Muslim world. 



Several studies focus on changes in application usage during the pandemic. A study reported the impact of COVID-19 outbreak on  E-learning for 16K students per day in 600 virtual classes in the Politecnico di Torino campus network \cite{e-learning} after the school closed. The authors testify how the Internet has proved robust to successfully cope with challenges while maintaining the university operations. ThousandEyes and App Annie \cite{thenewstack} report usage increase in gaming, VPN, and video while social apps show a reduced trend.


Two large Internet Exchange Points, i.e., Amsterdam Internet Exchange (AMS-IX) and Deutscher Commercial Internet Exchange (DE-CIX), reported general increase with exception of the CDN operator Akamai~\cite{akamai} which reported that US networks has not seen the same dramatic increase in Internet traffic as in Europe. AMS-IX reported 17\% general increase related to the COVID-19 crisis at the end of March 2020~\cite{ams-increase}, while DE-CIX reported 10\% general increase, 50\% increase in video-conferencing traffic (Skype, WebEx, Teams) and 25\% increase in online and cloud gaming and in traffic from the use of social media platforms~\cite{dec-increase}. Ookla speedtest is providing updated statistics on fixed and mobile network performance which is very useful to study individual performance statistics per country \cite{ookla}.



We complement and improve on previous work by presenting a comprehensive multi-perspective account of Internet performance during lockdown.
We examine the performance of the control and data planes as well as DNS and two distinct application level case studies. 

\label{sec:Related}

\section{Conclusions}
We have investigated the performance of the Internet during the first wave of covid-19 using a multi-perspective approach.
Our investigation examines BGP performance, intra-country latency, network outages, DNS stability and mobile network speed. 
We also assess the growth in video traffic using logs from a global video conferencing provider.  
Overall, we find that the Internet, across the world, has responded fairly well to the increasing traffic volumes generated by increased use that during the pandemic. We believe this is due to the robust architecture of the Internet, the scalable design of key components like DNS and BGP, clever engineering responses like the reduction of video definitions and long term investments in the Internet infrastructure.    


We also observe that several service providers, such as video and mobile network operators, have responded well to the increased load. Video services have seen growth of several hundred percent and even more for some customers and use of streaming services greatly increase outbound traffic. 
Traffic growth from before to during the pandemic has been significant, Very few other critical infrastructures would be able to sustain growth up to several hundred percent while still providing service with acceptable quality. 

\label{sec:Conclusion}

\bibliographystyle{ACM-Reference-Format}
\bibliography{ref}

\end{document}